\documentclass[12pt]{article}
\input epsf

\def\qqbar{\mbox{${\rm q\bar{q}}$}}

\newcommand{\mZ}{\ensuremath{m_{\rm Z}}}
\newcommand{\mH}{\ensuremath{m_{\rm H}}}
\newcommand{\Z}{\ensuremath{{\rm Z}}}

\newcommand{\gevcc}{\ensuremath{{\rm GeV}\!/c^2}}

\newcommand{\epem}{\ensuremath{{{\rm e}^+{\rm e}^-}}}
\newcommand{\epemto}{\ensuremath{{{\rm e}^+{\rm e}^- \to}}}

\newcommand{\glu}{\ensuremath{{\tilde{\rm g}}}}

\newcommand{\mglu}{\ensuremath{m_\glu}}

\newcommand{\eA}{\ensuremath{\varepsilon_1}}

\newcommand{\eC}{\ensuremath{\varepsilon_3}}
\newcommand{\eD}{\ensuremath{\varepsilon_{\rm b}}}

\newcommand{\eNP}{\ensuremath{\varepsilon_{\rm NP}}}
\newcommand{\eNPh}{\ensuremath{\varepsilon_{\rm NP}^{\rm had}}}
\newcommand{\eNPi}{\ensuremath{\varepsilon_{\rm NP}^{\rm inv}}}
\newcommand{\eNPl}{\ensuremath{\varepsilon_{\rm NP}^\ell}}
\newcommand{\xh}{\ensuremath{x_{\rm had}}}
\newcommand{\xl}{\ensuremath{x_\ell}}
\newcommand{\xinv}{\ensuremath{x_{\rm inv}}}

\def\Journal#1#2#3#4{{#1} {\bf #2} (#3) #4}
\def\etal{et al. }

\def\NPB{{\em Nucl. Phys.} {\bf  B}}

\def\PRL{\em Phys. Rev. Lett.}

\def\PRD{{\em Phys. Rev.} {\bf D}}
\def\ZPC{{\em Z. Phys.} {\bf C}}

\def\IJM{\em Int. J. Mod. Phys. {\bf A}}

\begin{document}

\pagenumbering{arabic}
\pagestyle{plain}

\date{}
\title{ \null\vspace{1cm}
The Light Gluino Mass Window Revisited
\vspace{1cm}}
\author{Patrick Janot\\
\footnotesize CERN, EP Division, CH-1211 Geneva 23, Switzerland\\
\footnotesize e-mail: Patrick.Janot@cern.ch}

\maketitle


\vspace{.2cm}
\begin{abstract}
\vspace{.2cm}
The precise measurements of the ``electroweak observables'' performed at 
LEP and SLC are well consistent with the standard model predictions. 
Deviations from the standard model arising from vacuum polarization diagrams 
(also called ``weak loop corrections'') have been constrained in 
a model-independent manner with the $\varepsilon$ formalism. Within 
the same formalism, additional deviations from new physics production 
processes can also be constrained, still in a model-independent way. 
For instance, a 95\%\,C.L. limit of 
$$ \Delta \Gamma_{\rm had} < 3.9\,{\rm MeV} $$ 
is set on the partial width of any purely hadronic exotic contribution 
to Z decays. When applied to the \epemto\ \qqbar\glu\glu\ process, it 
allows an absolute lower limit to be set on the gluino mass, 
$$ \mglu > 6.3\,\gevcc {\rm \ at\ 95\%\ C.L.}, $$
which definitely closes the so-called light gluino mass window.
\end{abstract}

\vfill

\eject

\section{Introduction}

The precise measurement of the Z total decay width at LEP  
and its agreement with its standard model prediction~\cite{LEPEWG} 
(with $\mH = 78^{+48}_{-31}$\,\gevcc)
\begin{equation}
\Gamma_{\rm Z}^{\rm exp} = (2495.2 \pm 2.3)\,{\rm MeV} \ \ \ {\rm and} \ \ \ 
\Gamma_{\rm Z}^{\rm SM} = (2495.9 \pm 2.4)\,{\rm MeV} 
\end{equation}
are often exploited
to constrain the cross section of new physics processes~\cite{ganis}. 
Indeed, under the assumption that new physics contributions
to the Z width are exclusively positive (as is the case
for processes kinematically allowed at $\sqrt{s} = \mZ$), 
this agreement allows a 95\% confidence level (C.L.) limit of 
\begin{equation}
\Delta \Gamma_\Z < 6.4\,{\rm MeV}
\end{equation}
to be set on any exotic contribution to $\Gamma_\Z$.

However, extensions of the standard model, such as supersymmetry
or technicolor, generate a whole set of new particles, which may or may 
not be produced in \epem\ collisions at $\sqrt{s} = \mZ$. The particles 
that are too heavy to be produced in Z decays may still contribute to the 
Z width through vacuum polarization diagrams with a generally undetermined 
sign. It may therefore well occur that negative contributions be sizeable 
and invalidate the widely used aforementioned limit on $\Delta \Gamma_\Z$.

It is the purpose of this letter to derive model-independent limits on 
additional contributions to Z decays, and to use these limits to 
unambiguously constrain the light gluino mass window. It is indeed
controversial if a light gluino \glu\ of mass below 5\,\gevcc\ is 
phenomenologically viable~\cite{farrar,clavelli}. A review of existing limits 
and of the related weak points can be found in Ref.~\cite{pdg}. In 
particular, a study of the QCD colour factors from four-jet angular 
correlations and the differential two-jet rate in Z decays, performed 
by ALEPH, allowed a 95\%\,C.L. lower limit of 6.3\,\gevcc\ to be set on 
\mglu~\cite{gunter}. However, it was argued by the light gluino 
defenders~\cite{farrar} that this limit was to be weakened because 
{\it (i)} the theory uncertainties were too aggressive; and {\it (ii)} 
no next-to-leading-order mass corrections were available for the four-jet 
angular correlations.

This letter is organized as follows. In Section~\ref{sec:vacuum},
the reader is reminded of the model-independent parametrization of the 
weak loop corrections to the electroweak observables according to the 
$\varepsilon$ formalism~\cite{altarelli}. This formalism is extended in 
Section~\ref{sec:hadronic} to the corrections caused by any new physics production 
process, in either the hadronic, the leptonic or the invisible final state. 
The result is applied to the $\epemto\ \qqbar\glu\glu$ process in 
Section~\ref{sec:gluino} and a model-independent lower limit on the gluino 
mass is obtained.

\section{Parametrizing the weak loop corrections}
\label{sec:vacuum}

Virtual contributions to the ``electroweak observables'' have been parametrized 
in a model-independent way by several authors. Here, the choice was 
made to parametrize the basic electroweak observables, {\it i.e.}, those 
sensitive to the weak loop corrections, with the (linearized) $\varepsilon$ 
formalism, according to~\cite{altarelli}

\begin{eqnarray}
\Gamma_{\rm Z} & = & \Gamma_{\rm Z}^0\, ( 1 + 1.35\eA -0.46\eC +0.35\eD ), \\
R_\ell & = & R_\ell^0\, (1 + 0.28\eA - 0.36\eC + 0.50\eD ), \\
\sigma_{\rm had} & = & \sigma_{\rm had}^0\, (1 - 0.03\eA +0.04\eC - 0.20\eD ), \\
g_V/g_A & = & \left(g_V/g_A\right)^0\, (1 + 17.6\eA - 22.9\eC ), \\
R_{\rm b} & = & R_{\rm b}^0\, (1 - 0.06\eA + 0.07\eC + 1.79\eD),
\end{eqnarray}
where
\begin{eqnarray}
\Gamma_{\rm Z}^0 & = & 2489.46\, (1+0.73\delta\alpha_S-0.35\delta\alpha)\,{\rm MeV},\\ 
R_\ell^0 & = & 20.8228\, (1+1.05\delta\alpha_S-0.28\delta\alpha), \\
\sigma_{\rm had}^0 & = & 41.420\, 
(1-0.41\delta\alpha_S+0.03\delta\alpha)\,{\rm nb},\\
\left(g_V/g_A\right)^0 & = & 0.075619 -1.32\delta\alpha, \\
R_{\rm b}^0 & = & 0.2182355,
\end{eqnarray}
are the Born approximations of the corresponding observables, {\it i.e.,}
without any weak loop corrections, and where the pure QCD- and QED-corrections
were parametrized as 
\begin{equation}
\delta\alpha_S = \frac{\alpha_S(\mZ) - 0.119}{\pi}\ \ {\rm and} \ \
\delta\alpha = \frac{\alpha(\mZ) - \frac{1}{128.90}}{\alpha(0)}.
\end{equation}

In the standard model, or in any theory that does not predict new open 
processes in \epem\ collisions at $\sqrt{s} = \mZ$, the three $\varepsilon$'s 
can then be fit to the precise measurements of LEP and SLC~\cite{LEPEWG}, 
summarized in Table~\ref{tab:data}. 
\begin{table}[htbp]
\begin{center}
\caption{\footnotesize Precise LEP and SLC measurements of the Z lineshape 
parameters ($\Gamma_{\rm Z}$, $R_\ell$, $\sigma_{\rm had}$), of $g_V/g_A$ 
and of $R_{\rm b}$, together with their correlation matrix. The last 
two measurements have been taken here as uncorrelated with the first 
three~\cite{bolek}.
\label{tab:data}}
\vspace{3mm}
\begin{tabular}{|l|l||rrrrr|} \hline\hline
Observable & Measurement & \multicolumn{5}{|c|}{Correlation matrix} \\ 
\hline\hline
$\Gamma_{\rm Z}$          &   $2495.2 \pm\ 2.4$\,MeV  & $1.000$ & & & &     \\ \hline
$R_\ell$            &   $20.767 \pm\ 0.025$     & $+0.004$ & $1.000$ & & & \\ \hline
$\sigma_{\rm had}$  &   $41.540 \pm\ 0.037$\,nb & $-0.297$ & $+0.183$ & $1.000$ & & \\ \hline
$g_V/g_A$           &   $0.07408 \pm\ 0.00068$  & $0.000$ & $0.000$ & $0.000$ & $1.000$ &    \\ \hline
$R_{\rm b}$         &   $0.21644 \pm\ 0.00065$  & $0.000$ & $0.000$ & $0.000$ & $0.000$ & $1.000$     \\ \hline\hline
\end{tabular}
\end{center}
\end{table}
\noindent
In this fit, the value of the strong and electromagnetic coupling constants 
were taken to be
\begin{equation} 
\alpha_S(\mZ) =  0.1183 \pm 0.0020~\cite{bethke} \ \ \ {\rm and} \ \ \ 
\alpha(\mZ)^{-1} = 128.95 \pm 0.05~\cite{LEPEWG}. \ \ \ 
\end{equation}
The validity of the latter is ensured in extensions of the standard model 
with only heavy new particles by the decoupling properties of QED, which 
allow the heavy particle contributions to be safely neglected in the running 
of $\alpha$ from 0 to \mZ. The value of $\alpha_S$ is well constrained by 
measurements performed directly at the Z resonance and does not suffer from 
this kind of uncertainties. 

The result of the fit, given in Table~\ref{tab:fitew}, is consistent with that
presented in Ref.~\cite{altarelli}, up to small deviations (less than 
$1\sigma$ or thereabout) caused by recent updates of the measurements and 
different variables included in the fit.
\begin{table}[htbp]
\begin{center}
\caption{\footnotesize Result of the fit of the $\varepsilon$'s to the precise
measurements of the five observables of Table~\ref{tab:data}. Also indicated, 
for comparison, is the result presented in Ref.~\cite{altarelli}
\label{tab:fitew}}
\vspace{3mm}
\begin{tabular}{|l|c|c|c|} \hline\hline
                      & $\eA\times 10^3$ & $\eC\times 10^3$ & $\eD\times 10^3$    
\\ \hline
This fit              & $5.4 \pm 1.0$ & $5.3 \pm 0.9$ & $-5.5 \pm 1.4$ 
\\ \hline\hline
Ref~\cite{altarelli} & $4.3 \pm 1.2$ & $4.5 \pm 1.1$ & $-3.8 \pm 1.9$ 
\\ \hline
\end{tabular}
\end{center}
\end{table}

\section{Model-independent limits on additional Z decays}
\label{sec:hadronic}

The fitted $\varepsilon$ values are usually interpreted in the standard
model to predict the value of the Higgs boson mass, or to constrain new
theories in which additional vacuum polarization diagrams would modify
the $\varepsilon$'s. Here, advantage is taken of the redundancy of the 
quantities in Eqs.~(3) to~(7) to set instead model-independent limits on 
additional Z decays, which would be caused by the existence of new 
particles light enough to be produced in \epem\ collisions at 
$\sqrt{s} = \mZ$.

For instance, such new particles could be produced and decay in such 
a way that they contribute only to hadronic Z decays (all quark flavours). 
Let \eNPh\ be the ratio of this new partial width $\Gamma_{\rm NP}$ to the 
total decay width of the Z without this new contribution. The first 
three observables are changed as follows,
\begin{eqnarray} 
\Gamma_{\rm Z} & \longrightarrow & \Gamma_{\rm Z}\,\left(1+1.00\eNPh\right), 
\ \ \ \ \ \left[ \Gamma_{\rm Z} + \Gamma_{\rm NP} \right] \\
R_\ell & \longrightarrow & R_\ell\,\left(1+1.43\eNPh\right), 
\ \ \ \ \ \left[ \left(\Gamma_{\rm had} + \Gamma_{\rm NP}\right)
/\Gamma_\ell \right] \\
\sigma_{\rm had} & \longrightarrow & \sigma_{\rm had}\, 
\left(1 - 0.57\eNPh\right),
\ \ \ \left[ \frac{12\pi}{m_{\rm Z}^2} 
 \frac{\Gamma_{\rm ee}(\Gamma_{\rm had} + \Gamma_{\rm NP})}
 {(\Gamma_{\rm Z} + \Gamma_{\rm NP})^2} \right]
\end{eqnarray}
while $(g_V/g_A)$ and $R_{\rm b}$ remain untouched. 
If the technical definition of the original $\varepsilon$'s is modified with 
respect to~\cite{altarelli} in such a way that they still only account for the
weak loop corrections, these changes modify in turn Eqs.~(3) to~(5) according 
to
\begin{eqnarray}
\Gamma_{\rm Z} & = & \Gamma_{\rm Z}^0\, ( 1 + 1.35\eA -0.46\eC +0.35\eD + 1.00\eNPh), \\
R_\ell & = & R_\ell^0\, (1 + 0.28\eA - 0.36\eC + 0.50\eD + 1.43\eNPh), \\
\sigma_{\rm had} & = & \sigma_{\rm had}^0\, (1 - 0.03\eA +0.04\eC - 0.20\eD - 0.57\eNPh),
\end{eqnarray}
and Eqs.~(6) and~(7) still apply. Similarly, a new physics contribution 
to the sole invisible decay width would modify the equations
according to 
\begin{eqnarray}
\Gamma_{\rm Z} & = & \Gamma_{\rm Z}^0\, ( 1 + 1.35\eA -0.46\eC +0.35\eD + 1.00\eNPi), \\
\sigma_{\rm had} & = & \sigma_{\rm had}^0\, (1 - 0.03\eA +0.04\eC - 0.20\eD - 2.00\eNPi),
\end{eqnarray}
and a new physics contribution to the leptonic decay width only (democratically 
in the three lepton flavours) to
\begin{eqnarray}
\Gamma_{\rm Z} & = & \Gamma_{\rm Z}^0\, ( 1 + 1.35\eA -0.46\eC +0.35\eD + 1.00\eNPl), \\
R_\ell & = & R_\ell^0\, (1 + 0.28\eA - 0.36\eC + 0.50\eD - 9.89\eNPl), \\
\sigma_{\rm had} & = & \sigma_{\rm had}^0\, (1 - 0.03\eA +0.04\eC - 0.20\eD + 
7.89\eNPl).
\end{eqnarray}
In each of the three cases, the new physics contribution \eNP\ can be fitted 
together with the other three $\varepsilon$'s to the five measured quantities.
The results of the three fits, all compatible with $\eNP = 0$, are listed in 
Table~\ref{tab:fitNP}.
\begin{table}[htbp]
\begin{center}
\caption{\footnotesize Results of the fits of the $\varepsilon$'s to the 
precise measurements of the five observables of Table~\ref{tab:data} when 
new physics Z decays are added, either in the hadronic, leptonic or
invisible final state. 
\label{tab:fitNP}}
\vspace{3mm}
\begin{tabular}{|l|c|c|c|c|} \hline\hline
Decay & $\eA\times 10^3$ & $\eC\times 10^3$ & $\eD\times 10^3$    
& $\eNP\times 10^3$  \\ \hline
Hadronic  & $5.7 \pm 1.0$ & $5.5 \pm 1.0$ & $-4.6 \pm 1.7$ & $-0.70 \pm 1.00$
\\ \hline
Leptonic  & $4.5 \pm 1.3$ & $4.6 \pm 1.1$ & $-4.1 \pm 1.6$ & $+0.13 \pm 0.11$
\\ \hline
Invisible & $5.4 \pm 1.0$ & $5.4 \pm 0.9$ & $-4.4 \pm 1.4$ & $-0.91 \pm 0.48$
\\ \hline\hline
\end{tabular}
\end{center}
\end{table}
Conservative upper limits on the \eNP's ({\it i.e}, on the new physics 
branching fractions) were derived at the 95\% confidence level by integrating 
their probability density functions in the physical region ($\eNP > 0$) only.
Indeed, $\eNP$ only accounts for real contributions to Z decays and can 
therefore only be positive. (All unknown virtual contributions are absorbed 
in the other $\varepsilon$'s.) 
These limits are reported in Table~\ref{tab:limit}, together with the 
corresponding limits on the new physics partial width and on the new physics 
cross section at the Z peak. For completeness, $3\sigma$- and $5\sigma$-limits 
are also indicated.
\begin{table}[htbp]
\begin{center}
\caption{\footnotesize Limits at 95\%\,C.L. on the new physics branching ratio,
partial width and cross section at the Z peak, in the hadronic, the leptonic
and the invisible final states. The limits at $3\sigma$ (99.63\%\,C.L.) and 
$5\sigma$ (99.99994\%\,C.L.) are also indicated. 
\label{tab:limit}}
\vspace{3mm}
\begin{tabular}{|l|c|c|c|c|} \hline\hline
Final state & Hadronic & Leptonic & Invisible \\ \hline\hline
$\Delta{\rm BR}_{95}$ ($10^{-3}$)      & 1.56 & 0.31 & 0.54 \\ \hline
$\Delta\Gamma_{95}$ (MeV)              & 3.9  & 0.77 & 1.33 \\ \hline
$\Delta\sigma_{95}$ (pb)               & 66.9 & 13.2 & 22.9 \\ \hline\hline
$\Delta{\rm BR}_{3\sigma}$ ($10^{-3}$) & 2.52 & 0.43 & 0.93 \\ \hline
$\Delta\Gamma_{3\sigma}$ (MeV)         & 6.3  & 1.07 & 2.32 \\ \hline
$\Delta\sigma_{3\sigma}$ (pb)          & 108. & 18.3 & 39.8 \\ \hline\hline
$\Delta{\rm BR}_{5\sigma}$ ($10^{-3}$) & 4.45 & 0.65 & 1.78 \\ \hline
$\Delta\Gamma_{5\sigma}$ (MeV)         & 11.1 & 1.61 & 4.45 \\ \hline
$\Delta\sigma_{5\sigma}$ (pb)          & 191. & 27.5 & 76.5 \\ \hline\hline
\end{tabular}
\end{center}
\end{table}

\begin{figure}[htbp]
\begin{picture}(160,80)
\put(71,70){(a)}
\put(150,70){(b)}
\put(0,-2){\epsfxsize85mm\epsfbox{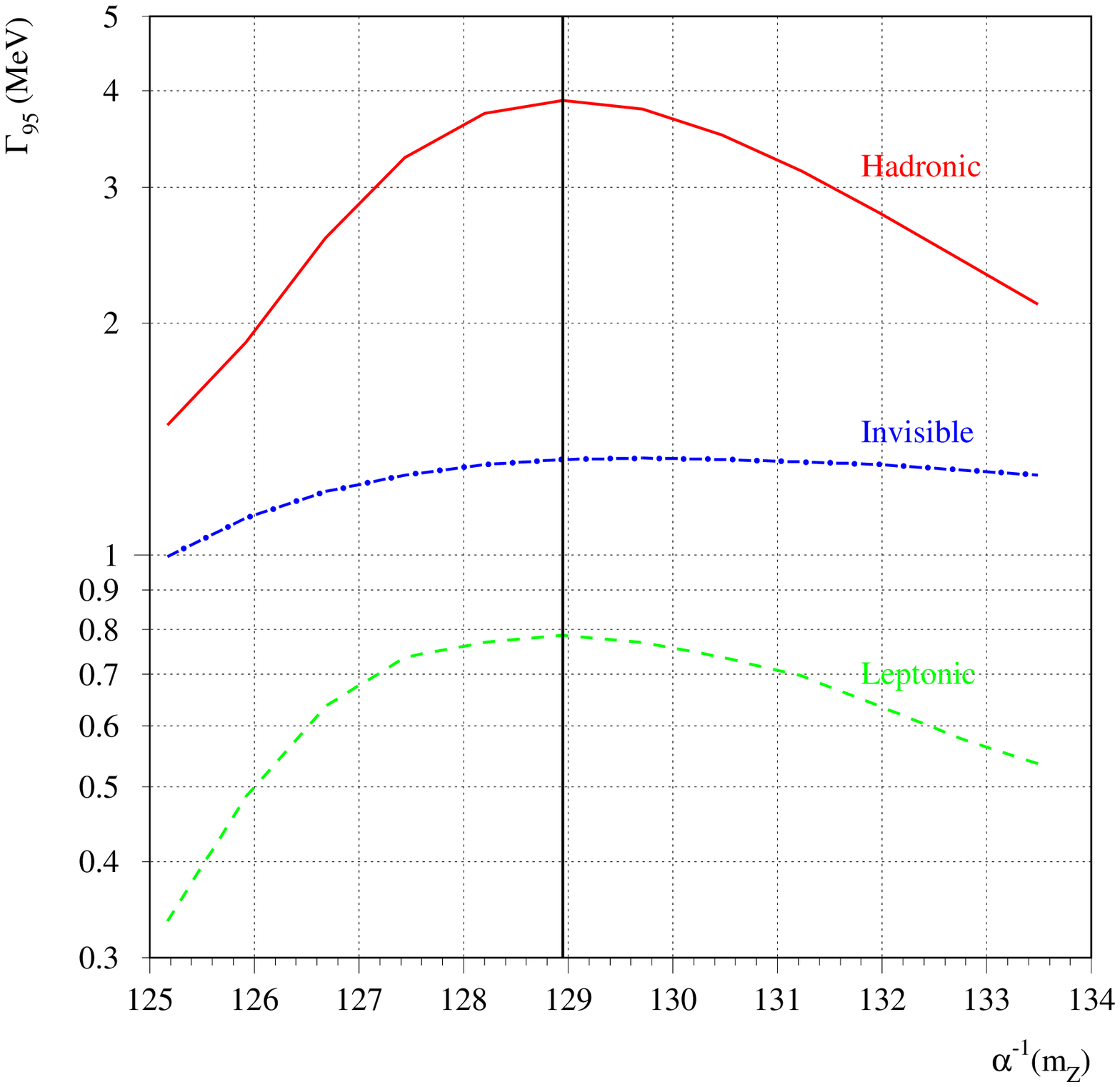}}
\put(80,-2){\epsfxsize85mm\epsfbox{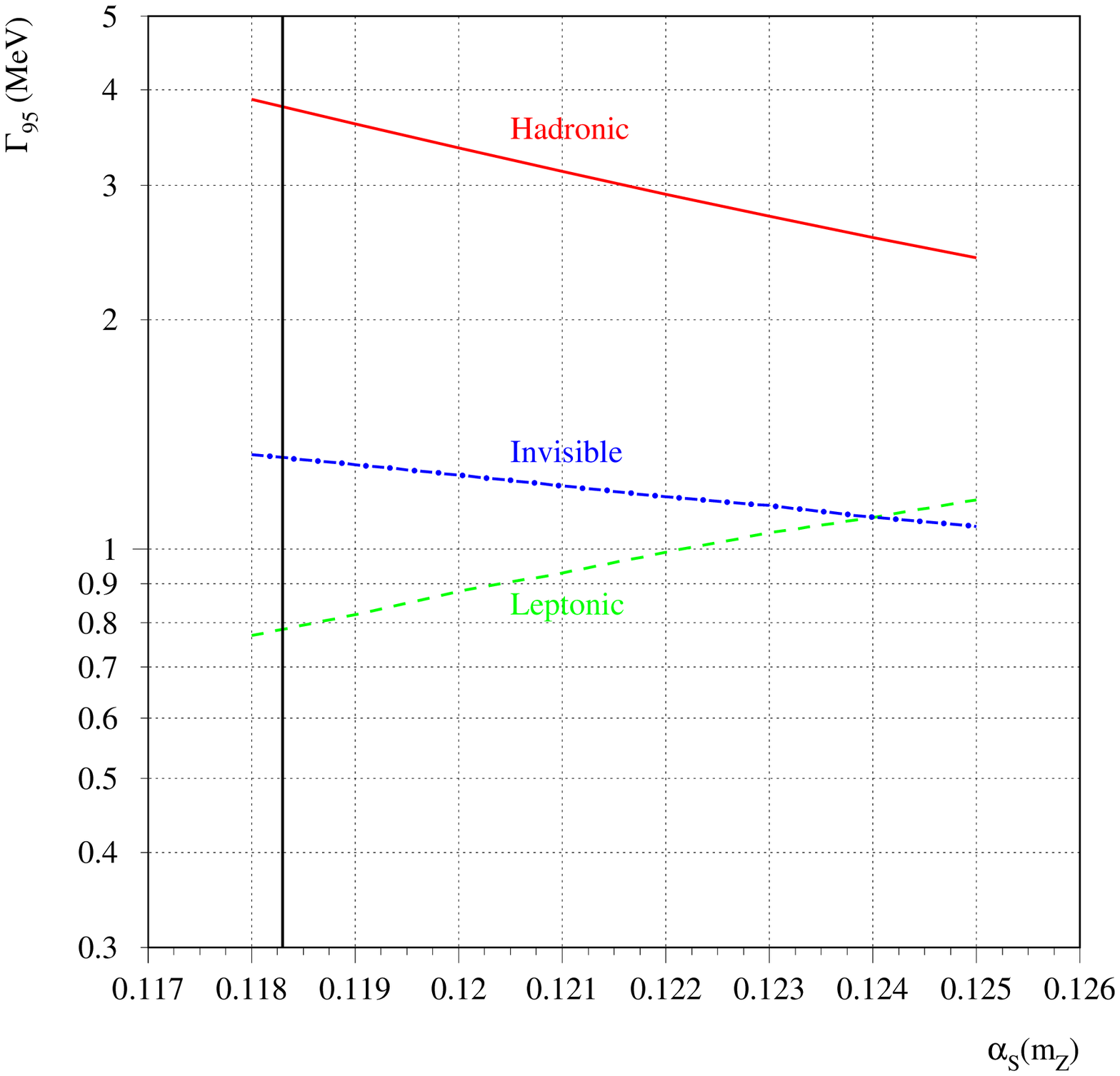}}
\end{picture}
\caption[ ]
{\protect\footnotesize The 95\%\,C.L. upper limits on the new physics partial 
width, in the hadronic (full curve), the leptonic (dashed curve) and 
the invisible (dot-dashed curve) as a function of (a) $1/\alpha(\mZ)$
and (b) $\alpha_S(\mZ)$. The vertical lines indicate the standard values
of the coupling constants, chosen to perform the fits. 
\label{fig:alpha}} 
\end{figure}

Because the existence of new particles could also modify the 
evolution of the electromagnetic coupling constant from 0 to \mZ, it 
was checked whether these limits could be affected by a variation of 
$\alpha(\mZ)$. This check is illustrated in Fig.~\ref{fig:alpha}a, where
it appears that the standard model value of $\alpha$ yields the largest
upper limits on the three partial widths. The limits of Table~\ref{tab:limit}
are therefore conservative in this respect. 

New light hadronic flavours would also modify the evolution of the 
strong coupling constant. Although, as already mentioned, $\alpha_S$
is constrained by measurements performed directly at the Z mass scale, 
it is also determined with at least as accurate low energy measurements 
extrapolated at \mZ. A different scaling law for the latter would 
modify the world average of $\alpha_S$. However, a contribution from 
new coloured scalars or fermions would always slow down the 
running from low energy to \mZ\ so as to increase the value 
of $\alpha_S(\mZ)$~\cite{running}. As shown in Fig.~\ref{fig:alpha}b, 
such an increase of $\alpha_S$ would render more constraining
the limits on the hadronic and invisible partial widths. The limit
on the leptonic width would be slightly weakened, but would anyway remain 
the strongest of the three constraints. The conservative choice of 
ignoring this effect was made throughout.

Finally, it may be argued that a new hadronic contribution to Z decays
would decrease the value of $\alpha_S$ fitted from the Z lineshape. 
However, the $\alpha_S$ world average and its uncertainty (saturated by 
common theory errors), and therefore the result of the present fit, do 
not change noticeably when the $\alpha_S$ measurement from the Z 
lineshape in taken out.

\begin{figure}[htbp]
\begin{picture}(160,80)
\put(0,-2){\epsfxsize85mm\epsfbox{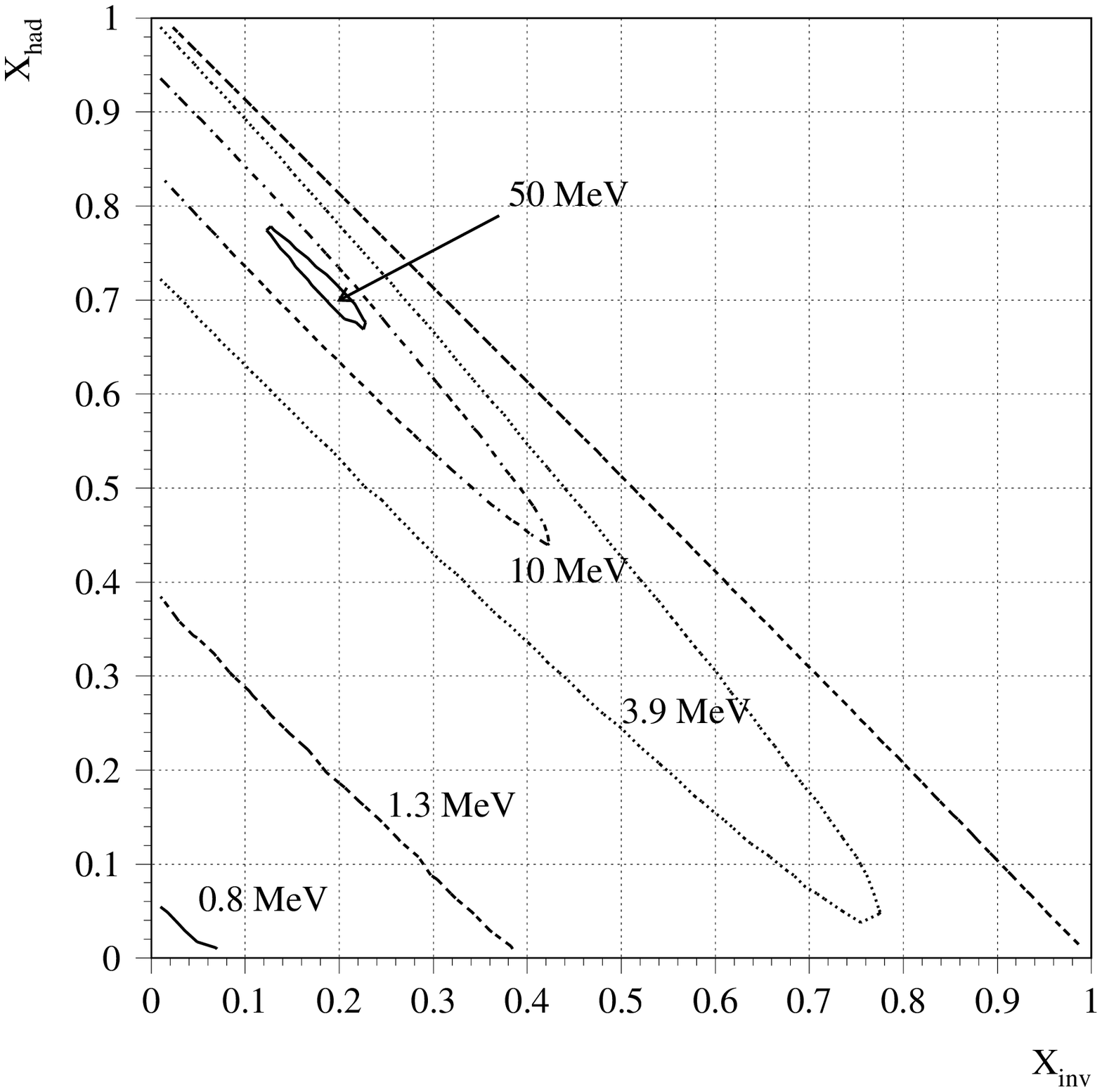}}
\put(71,70){(a)}
\put(150,70){(b)}
\put(80,-2){\epsfxsize85mm\epsfbox{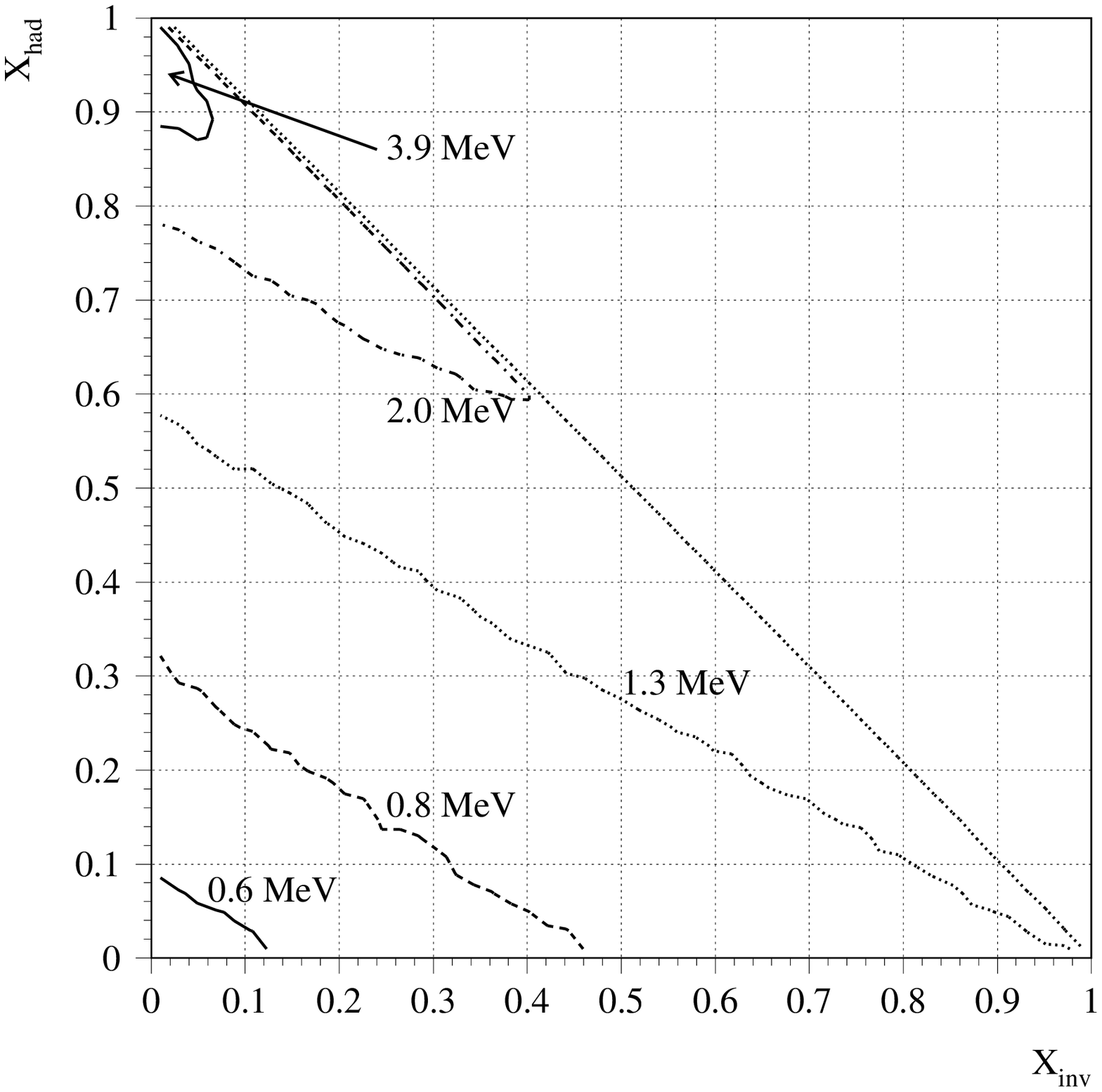}}
\end{picture}
\caption[ ]
{\protect\footnotesize The 95\%\,C.L. upper limit on the new physics 
contribution to the total decay width, as a function of the 
fraction of hadronic and invisible final states arising from   
the new particle production, with (a) leptonic decays democratic 
in the three flavours and (b) leptonic decays in only one 
flavour ($\mu$ or $\tau$). The contours indicate the values
of $x_{\rm inv}$ and $x_{\rm had}$ for which the limit amounts
to 0.6, 0.8, 1.3, 2.0, 3.9, 10 or 50 MeV.
\label{fig:plane}}
\end{figure}

The fit of \eNP\ can be repeated in any other configuration of hadronic, 
leptonic and invisible contributions to the Z decays from the new physics 
process. Let \xh, \xl\ and \xinv\ be the fractions of hadronic, leptonic 
and invisible final states produced by the new physics process under 
consideration. By definition, a final state which is neither hadronic nor 
leptonic is called invisible, therefore $\xinv + \xh + \xl = 1$. The Z 
lineshape parameters of Eqs.~(3) to~(5) are modified according to 
\eject
\begin{eqnarray} 
\Gamma_{\rm Z} & \longrightarrow & \Gamma_{\rm Z}\,\left[1+\eNP\right], \\
R_\ell & \longrightarrow & R_\ell\,
\left[1+\eNP \left (1.43\xh -9.89\xl\right)\right], \\
\sigma_{\rm had}& \longrightarrow & \sigma_{\rm had}\, 
\left[1 + \eNP \left(-2.00+1.43\xh+9.89\xl)\right) \right].
\end{eqnarray}
The  (\xh, \xinv) plane was scanned and the fit performed at
each point, yielding a limit on \eNP\ everywhere in this plane. 
The corresponding 95\%\,C.L. limit on $\Delta\Gamma_{\rm Z}$, 
the new physics contribution to the Z total width, is 
displayed in Fig.~\ref{fig:plane}a. Similarly, the 
limit on $\Delta\Gamma_{\rm Z}$ under the hypothesis that the 
new particle production contributes to hadronic and invisible 
final states, and to only one lepton flavour ($\mu$ or $\tau$), 
is shown in Fig.~\ref{fig:plane}b.

Very constraining limits on $\Delta\Gamma_{\rm Z}$ are set all over 
the plane, but no absolute limit can be obtained when the new 
particle production leads to fractions in the hadronic, invisible 
and leptonic (three flavours) final states identical to the Z 
branching fractions. In this case, only $\Gamma_{\rm Z}$ depends 
on \eNP. Equations~(4) to~(7) no longer yield an independent 
determination of \eA\ and \eC\ with meaningful accuracy, because they all 
depend on the same linear combination of the two quantities. As 
a result, the new particle contribution to the Z width can always 
be cancelled by the $(1.35\eA - 0.46\eC)$ virtual contribution,  
if a sufficient amount of fine tuning takes place. 

Whether or not 
this amount of fine tuning is acceptable would need different 
measurements and/or more theory to decide.

\section{A model-independent limit on the gluino mass}
\label{sec:gluino}

\begin{figure}[htbp]
\begin{picture}(160,95)
\put(25,-2){\epsfxsize110mm\epsfbox{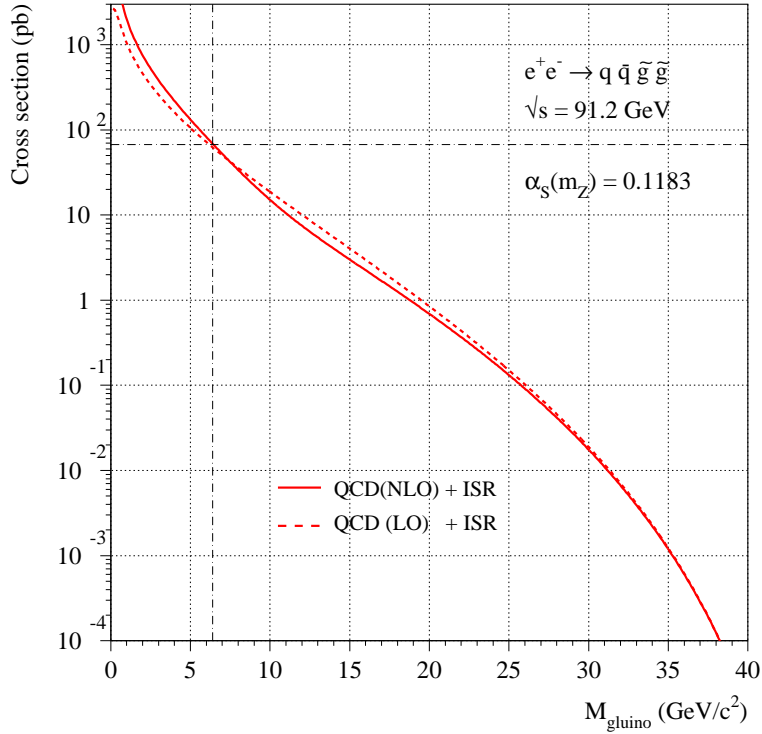}}
\end{picture}
\caption[ ]
{\protect\footnotesize The production cross section of the process 
\epemto\ \qqbar\glu\glu\ as a function of the gluino mass, at 
$\sqrt{s} = \mZ$ (dashed curve: QCD leading order, full curve: 
QCD next-to-leading order). Also indicated is the 95\%\,C.L. upper 
limit on this cross section from precise LEP and SLC measurements 
and  the corresponding upper limit on the gluino mass (dot-dashed 
lines).
\label{fig:gunion}} 
\end{figure}

The results obtained in Section~\ref{sec:hadronic} can be applied
to a variety of new processes. In this letter, they are used to 
constrain the cross section of the gluino production at 
$\sqrt{s} = \mZ$ in the process
\begin{equation}
\epemto\ \qqbar\glu\glu,
\end{equation}
displayed in Fig.~\ref{fig:gunion} as a function of the gluino 
mass~\cite{gunion}. Because this cross section is the product of the 
$\epemto\ \qqbar$ cross section and a term that describes the gluon 
splitting into a gluino pair, the uncertainties related to the  weak 
loop corrections to $\epemto\ \qqbar\glu\glu$ were alleviated by using 
the measured \qqbar\ cross section in the prediction.

When the gluino is light, the final state arising from this process is 
purely hadronic irrespective of the gluino decay and hadronization, and 
therefore contributes solely to the Z hadronic decay width in all quark 
flavours. A 95\%\,C.L. upper limit on the production cross section at 
$\sqrt{s} = \mZ$ can then be set at 67\,pb (Table~\ref{tab:limit}). The 
corresponding lower limit on the gluino mass can be read off 
from the curve in Fig.~\ref{fig:gunion}, and is (including a 2\% systematic
uncertainty on $\alpha_S(\mZ)$)
\begin{eqnarray}
\mglu & > & 6.3\,\gevcc {\rm \ at\ 95\%\ C.L.}, \\
\mglu & > & 5.3\,\gevcc \ (3\sigma\ {\rm limit}),\\
\mglu & > & 4.2\,\gevcc \ (5\sigma\ {\rm limit}).
\end{eqnarray}

\parskip 1.6mm

This limit confirms the one obtained by ALEPH~\cite{gunter} with an
independent study of the QCD colour factors, which makes no use
of the absolute Z decay rates, and that derived from the running of 
the strong coupling constant~\cite{fodor}, which checks in addition 
the compatibility of the $\alpha_S$ measurements at all energy scales. 
Because it would add fully independent information and because it would 
avoid the conservative choices of Section~\ref{sec:hadronic} to be made, 
a combination of these results would further consolidate the light gluino 
exclusion. For instance, a combination of the present limit and that of
Ref.~\cite{gunter} yields lower limits on \mglu\ of 6.8\,\gevcc\ and 
5.7\,\gevcc\ at 95\% and 99.63\%\,C.L., respectively.


The present limit is not affected by the criticisms put forward 
to invalidate the previous results. 
It is only if three different fine tuning processes took 
place, {\it i.e.,} 
\begin{enumerate}
\item if other new particles were produced in association with the gluino
with a cross section of the order of or larger than that of the gluino 
production, but still were not directly detected at LEP;
\item if these processes led to final states such that the overall fractions 
of hadronic, invisible and leptonic new decays be similar to those of 
the Z decays, for all lepton and quark flavours;
\item and if additional new physics yielded large negative virtual contributions 
to the Z total decay width (from the $1.35\eA - 0.46\eC$ combination) to 
exactly compensate this multiple new particle 
production; 
\end{enumerate}
that the limit derived with the method presented in this letter would 
not hold. 
I leave it to the champions of the light gluino scenario to find a theory 
in which this devilish conspiracy could take place.

\section{Conclusion}

A method to derive model-independent limits on new physics contributions
to Z decays has been presented. No general upper limit on the total Z 
decay width could be obtained, but very stringent constraints apply when 
the final states produced by the new physics process of interest are 
known. In particular, conservative upper limits have been put on 
$\Gamma_{\rm Z}$ of 0.55, 1.3 and 3.9 MeV in the case of purely 
leptonic ($\mu$ or $\tau$), invisible and hadronic final states. 

When applied to the $\epemto\ \Z \to \qqbar\glu\glu$ process, it allows
a model-independent lower limit to be set on the gluino mass:
$$ \mglu > 6.3\,\gevcc {\rm \ at\ 95\%\,C.L.} $$
The light gluino mass window is closed.

\subsection*{Acknowledgments}

This work benefited from fruitful discussions with and useful comments from 
G.\,Altarelli, P.\,Azzuri, J.-B.\,De Vivie, G.\,Dissertori, G.\,Ganis, 
J.-F.\,Grivaz, B.\,Pietrzyk, G.\,Sguazzoni and R.\,Tenchini.

\end{document}